\documentclass[reprint,amsmath,amssymb,aps,prb,superscriptaddress]{revtex4-1}  % for review and submission
\usepackage{graphicx}  % needed for figures
\usepackage{dcolumn}   % needed for some tables
\usepackage{bm}        % for math
\usepackage{amssymb}   % for math
\usepackage{color}

\begin{document}

\preprint{APS/123-QED}

\title{Strain broadening of the 1042-nm zero-phonon line of the NV$^-$ center in diamond: a promising spectroscopic tool for defect tomography}
\author{T.B.\,Biktagirov}
\affiliation{Kazan Federal University, Kazan, 420008 Russia}
\author{A.N.\,Smirnov}
\author{V.Yu.\,Davydov}
\affiliation{Ioffe Institute, St. Petersburg, 194021 Russia}
\author{M.W.\,Doherty}
\affiliation{Laser Physics Centre, Research School of Physics and Engineering, Australian National University,
Australian Capital Territory 2601, Australia}
\author{A.\,Alkauskas}
\affiliation{Center for Physical Sciences and Technology, Vilnius LT-10257, Lithuania}
\affiliation{Department of Physics, Kaunas University of Technology, Kaunas LT-51368, Lithuania}
\author{B.C.\,Gibson}
\affiliation{ARC Centre of Excellence for Nanoscale BioPhotonics, School of Science, RMIT University, Melbourne,
Victoria 3001, Australia}
\author{V.A.\,Soltamov}
\affiliation{Kazan Federal University, Kazan, 420008 Russia}
\affiliation{Ioffe Institute, St. Petersburg, 194021 Russia}

\date{\today}%

\begin{abstract}

The negatively charged nitrogen-vacancy (NV$^-$) center in diamond is a promising candidate for many quantum applications. Here, we examine the splitting and broadening of the center's infrared (IR) zero-phonon line (ZPL). We develop a model for these effects that accounts for the strain induced by photo-dependent microscopic distributions of defects. We apply this model to interpret observed variations of the IR ZPL shape with temperature and photoexcitation conditions. We identify an anomalous temperature dependent broadening mechanism and that defects other than the substitutional nitrogen center significantly contribute to strain broadening. The former conclusion suggests the presence of a strong Jahn-Teller effect in the center's singlet levels and the latter indicates that major sources of broadening are yet to be identified. These conclusions have important implications for the understanding of the center and the engineering of diamond quantum devices. Finally, we propose that the IR ZPL can be used as a sensitive spectroscopic tool for probing microscopic strain fields and performing defect tomography.
\begin{description}
\item[PACS numbers]
61.72.Hh, 71.55.-i, 76.70.Hb, 61.72.jd
\end{description}
\end{abstract}

\pacs{Valid PACS appear here}% PACS, the Physics and Astronomy
\maketitle

\section{\label{sec:level1}Introduction%\protect\\ %The line
%break was forced \lowercase{via} \textbackslash\textbackslash
}
In recent years, the negatively charged nitrogen-vacancy (NV$^-$) center in diamond has attracted numerous research efforts aimed towards a wide range of applications. This color center exhibits long ground-state spin coherence and optically induced spin polarization over a broad range of wavelengths in ambient conditions, which makes it a perfect candidate for quantum sensing applications,\cite{qs1,qs2,qs3,qs4,qs5,qs6} room temperature operated masers,\cite{maser} and quantum computing.\cite{qc1,qc2,qc3}

The inhomogeneous broadening of the optical and spin resonances of NV$^-$ centers is a key limitation of many of these applications. It is currently believed that this inhomogeneous broadening is principally caused by the inhomogeneous strain and electromagnetic fields generated by random microscopic distributions of defects in the diamond, such as the substitutional nitrogen (N) center.\cite{stoneham69,davies70,davies71} However, the correlation between microscopic defect density and inhomogeneous broadening has not yet been quantitatively studied, and so this hypothesis has not yet been rigorously tested. Better understanding of the sources of inhomogeneous broadening will ultimately support superior engineering of diamond quantum devices.

\begin{figure*}
    \includegraphics{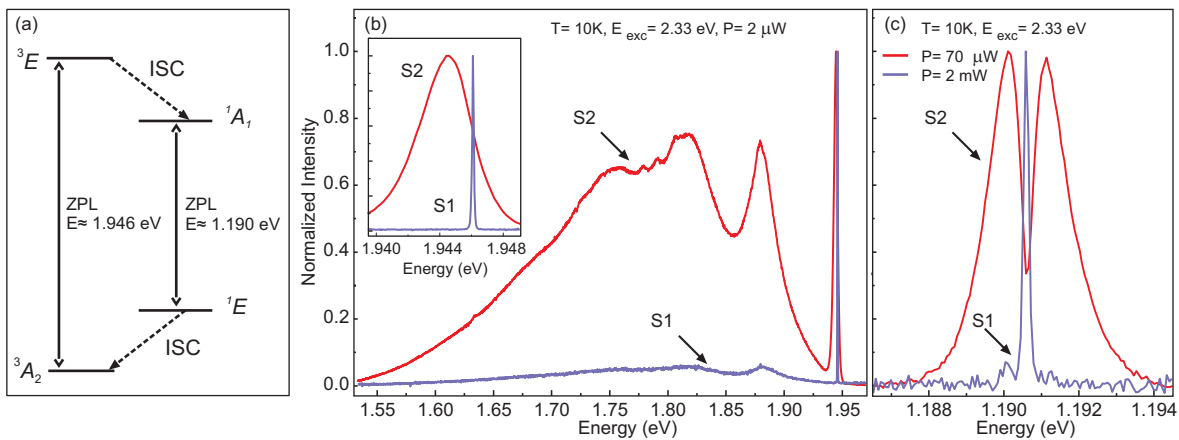}
\caption{\label{fig1} (a) The NV$^-$ center's electronic structure, including the low temperature visible E~$\approx$~1.946~eV and infrared E~$\approx$~1.190~eV ZPL energies indicated by the solid arrows, and the spin-dependent non-radiative intersystem crossings (ISCs) between the triplet and singlet levels. (b,c) PL spectra recorded at T=10~K in samples S1 and S2 in the ranges of the NV$^-$ center's visible and IR ZPLs, respectively. The energy (E$_\mathrm{exc}$) and power (P) of the excitation laser are as indicated. The inset in (b) shows the visible ZPL with enlarged scale.}
\end{figure*}

The NV$^-$ center manifests two zero-phonon lines (ZPL) [Fig.1 (a)], the visible 1.946~eV (637~nm) and infrared (IR) 1.190~eV (1042~nm) associated with triplet-triplet ${^3}\textit{A}{_2}\leftrightarrow{^3}\textit{E}$ and singlet-singlet ${^1}\textit{E}\leftrightarrow{^1}\textit{A}{_1}$ transitions, respectively.\cite{Doherty_PhysRep_2013} The ${^1}\textit{E}\leftrightarrow{^1}\textit{A}{_1}$ transition plays a crucial role in the spin dynamics of the NV$^-$ center because, along with the triplet-singlet non-radiative intersystem crossings (ISCs) [shown by the dashed arrows in Fig.~1(a)], it is part of the spin-dependent channel that enables the optical spin polarization and readout of the center's ground state spin.\cite{Doherty_PhysRep_2013} Thus, further study of the IR ZPL is important to the understanding of the properties of the singlet levels and the ISCs.\cite{goldman1,goldman2}

In this paper we report a detailed study of the splitting and inhomogeneous broadening of the NV$^-$ center's IR ZPL. We interestingly find that the splitting and broadening can be reduced by increasing excitation intensity or reducing excitation wavelength. To explain these observations, we develop a model for the inhomogeneous broadening that includes the strain induced by photo-dependent microscopic distributions of point defects. We use this model to distinguish strain broadening from other sources of broadening and, by doing so, identify an anomalous temperature dependent broadening mechanism, which potentially provides additional evidence of a strong Jahn-Teller effect in the $^1$E level. We show that the quantified strain broadening cannot be completely accounted for by the presence of N centers, thereby suggesting that there are major sources of broadening yet to be identified. We propose that the IR ZPL has the potential to be used as a sensitive spectroscopic tool to probe the microscopic distribution of strain fields and, with further work that identifies the major sources of strain, to perform tomography of defect concentrations.

\section{\label{sec:level1}Experimental Method}

Two different diamond samples containing NV$^-$ centers were studied in this work. The main difference between the samples was the concentration of N centers. Sample 1 (S1) was a type IIa single crystal diamond plate with dimensions 3$\times$3$\times$0.3~mm that was fabricated commercially by Element-6 using chemical-vapor deposition (CVD). The initial concentration of N was $\lesssim$100~ppb. The sample was irradiated to a total dose of 10$^{18}$~cm$^{-2}$ with 900~keV  electrons at room temperature. Sample~2 (S2) was a type Ib single crystal diamond plate with dimensions 3$\times$3$\times$0.3~mm that was fabricated commercially by Element-6 using high-pressure high-temperature (HPHT) growth and initially contained $\lesssim$200~ppm of N. The sample was irradiated with 3~-~5 MeV neutrons at temperatures close to room temperature to a total dose of 10$^{18}$~cm$^{-2}$. S2 was then annealed at a temperature \textit{T}~=~800$^\circ$C for two hours in the presence of a forming gas (H2) in order to achieve a high density of NV$^-$ centers via the capture of mobile vacancies by N centers.\cite{Mita_1996,Budker_PRB_NV_production_2010} S1 was not subjected to any thermal treatment. Thus, in summary, S1 has low densities of NV and N, whereas S2 has high densities.

The optical properties of the NV$^-$ centers in each sample was studied by performing micro-photoluminescence (PL) measurements using a Horiba Jobin-Yvon T64000 spectrometer equipped with a confocal microscope and a closed-cycle helium system for cryogenic microscopy. A Nd:YAG-laser line 2.33~eV (532~nm) and a He-Ne-laser line 1.96~eV (632.8~nm) were used for photoexcitation. The beam imaging, normal to the surface, was focused by a Mitutoyo 100xNIR (NA~=~0.50) objective lens into each sample with a spot size of $\sim$2~$\mu$m. The same objective collected the PL. We used a CCD camera together with 1800 lines/mm and 600 lines/mm gratings to measure the PL spectra with the spectral resolution of $\approx$0.05~meV in the visible and IR ranges, respectively.

\section{\label{sec:level1}Results and Analysis}

We start with the comparison of the PL spectra of the NV$^-$ centers in samples S1 and S2 that are presented in Fig.~1(b,c). At a glance, there are significant differences in both the visible and IR ZPLs of the samples.
In S1, narrow ZPLs were observed with simple Gaussian inhomogeneous lineshapes characterized by full widths at half maximum (FWHM) of $\sim$0.14 meV (visible) and $\sim$0.2 meV (IR). In S2, the visible and IR ZPLs differ significantly. The visible ZPL in S2 has shifted and broadened relative to S1 [inset in Fig 1(b)], whereas the IR ZPL in S2 has instead split and broadened.

The differing inhomogeneous lineshapes of the visible and IR ZPLs in S2 can be qualitatively explained by recalling that the ZPLs have different susceptibilities\cite{Rogers_Doherty_2015,Doherty_NJP_2011} to inhomogeneous strain and electric fields induced by random distributions of defects.\cite{stoneham69,davies70,davies71} Being a type Ib diamond, the most prevalent defect in S2 is expected to be the N center. In thermal equilibrium, the N center adopts a neutral charge state N$^0$. In this state, the center is paramagnetic and spontaneously undergoes a large distortion where one of the N-C bonds is substantially elongated.\cite{Loubser_vanWyk_1978} The known stress susceptibility parameters of the visible and IR ZPLs\cite{Rogers_Doherty_2015} imply that the strain induced by a N$^0$ center's distortion will primarily shift the visible ZPL of a proximal NV$^-$ center, whilst primarily splitting its IR ZPL. Averaging over the strain fields induced by many N$^0$ centers as well as over many NV$^-$ sites, one therefore expects to observe an inhomogeneously broadened visible ZPL without a splitting and an inhomogeneously broadened IR ZPL with a discernable splitting, as exhibited in Fig. 1(b,c).

\begin{figure}[!h]%
\begin{center}
   \includegraphics[width=0.95\columnwidth]{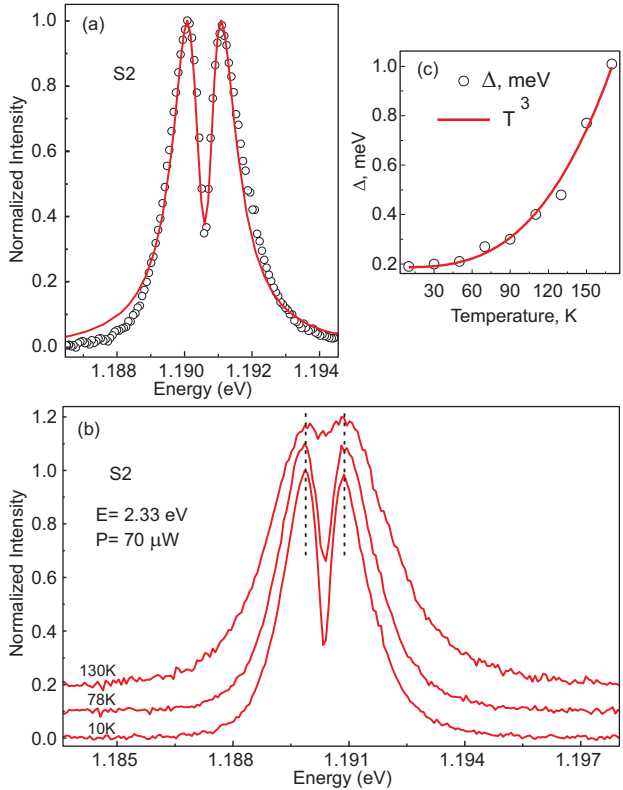}
\caption{\label{fig2} (a) The fit (solid line) of the IR ZPL spectra (circles) of S2 from Fig. 1(c) using the inhomogeneous lineshape expression (\ref{convolution}). The fit parameters are $\gamma$~=~1.7$\times$10$^{-4}$ and $\Delta$~=~0.18 meV. (b) Temperature dependence of the IR ZPL shape for S2. The dashed lines serve as a guide to demonstrate the temperature independence of the strain-induced splitting. The fixed photoexcitation energy and power are denoted in the figure. (c) Plot showing the temperature dependence of the fitted Gaussian width $\Delta$. Experimental (circles) and fit (solid line) obtained using the power law $\Delta$(\textit{T})~=~$\Delta_0$+$\textit{a}\times\textit{T}~^3$. The fit parameters are $\Delta_0$~=~(0.18$\pm$0.1)~meV and $\textit{a}$~=~1.65$\times10^{-4}$~meV~K$^{-3}$. }
\end{center}
\end{figure}

Now, under photoexcitation with photon energy $\gtrsim1.7$ eV,\cite{Farrer_1969,Isberg_PRB_2006} the N$^0$ center can be photoionized to form the positive charge state N$^+$. In this state, the center is non-paramagnetic and returns to an undistorted configuration where the N atom resides at the substitutional site.\cite{stoneham92} As a result, N$^+$ generates an electric field due to its charge, but induces minimal strain. Since first-principles theory predicts that the IR ZPL is not susceptible to electric fields,\cite{Doherty_NJP_2011} the photoionization of N$^0$ to N$^+$ should lead to narrowing of the IR ZPL, thereby providing a means to test that N$^0$ is indeed the dominant source of inhomogeneous broadening. The same is not true for the visible ZPL because it is known to be susceptible to electric fields.\cite{Doherty_PhysRep_2013} Therefore, any narrowing of the visible ZPL due to the smaller strain generated by N$^+$ will be countered by additional broadening due to the electric field it produces. Consequently, in the following we focus on the IR ZPL as a tool to probe the sources of inhomogeneous broadening.

To quantitatively describe the inhomogeneous broadening of the IR ZPL, we adapt the model of defect-induced strain broadening proposed in Refs \onlinecite{Klimin_Malkin_2010} and \onlinecite{Malkin_PRB_2012} for rare earth doped inorganic crystals. The key assumptions of model are that the strain contributions from individual point defects add linearly and the crystal is well approximated by an isotropic elastic continuum. In the model, the strain-induced changes of the two transition energies $\nu_\pm$ of the IR ZPL are
\begin{equation}
\label{1}
\begin{array}{c}
 \displaystyle
\nu_\pm  = V(A_1)e(A_1) + V(A_1^{'})e_1(A^{'}) \pm
\\
\\
 \displaystyle
\sqrt{V^2(E)[e_1(E)+e_2(E)]^2 + V^2(E^{'})[e_1(E^{'}) + e_2(E^{'})]^2}
\:.
\end{array}
\end{equation}
where $\textit{e}$(\textit{A}$_1$)~=~\textit{e}$_{xx}$ + \textit{e}$_{yy}$ + \textit{e}$_{zz}$, \textit{e}(\textit{A}$_1$$^{'}$) = $\frac{1}{2}$(2\textit{e}$_{zz}$ - \textit{e}$_{xx}$ - \textit{e}$_{yy}$), \textit{e}${_1}$(\textit{E}) = \textit{e}$_{yy}$~-~\textit{e}$_{xx}$, \textit{e}${_2}$(\textit{E})~=~2\textit{e}$_{xy}$, \textit{e}${_1}$(\textit{E}$^{'}$)~=~2\textit{e}$_{xz}$ and \textit{e}${_2}$(\textit{E}$^{'}$)~=~2\textit{e}$_{yz}$ are the deformation tensor components that have been defined to have explicit C$_{3v}$ symmetry properties in the coordinate frame of the NV center, and \textit{V}(\textit{A}$_1$), \textit{V}(\textit{A}$_1$$^{'}$), \textit{V}(\textit{E}), \textit{V}(\textit{E}$^{'}$) are the corresponding strain susceptibility parameters of the IR ZPL. The susceptibility parameters can be obtained from those reported in terms of stress in the crystal coordinate system in Ref \onlinecite{Rogers_Doherty_2015} by applying the elastic constants of the crystal and performing the required coordinate frame rotations. One finds that \textit{V}(\textit{A}$_1$)=~636.48, \textit{V}(\textit{A}$_1$$^{'}$)=~--437.76, \textit{V}(\textit{E})=~--1567.17, \textit{V}(\textit{E}$^{'}$)=~--1373.22 in units of meV/strain for the IR ZPL. Note that from the above expression for $\nu_\pm$, it can be seen that the IR ZPL splits due to the deformations of \textit{E} symmetry, whilst it shifts due to those of \textit{A}$_1$ symmetry.

Following Ref. \onlinecite{Malkin_PRB_2012}, assuming locally uniform densities of point defects, the statistical distribution function of the deformation tensor components \textbf{e}~$\equiv$~[\textit{e}$_{1}$(\textit{E}), \textit{e}$_{2}$(\textit{E}), \textit{e}$_{1}($\textit{E}$^{'}$), \textit{e}$_{2}$(\textit{E}$^{'}$)] is
\begin{equation}
\label{11}
g(\textbf{e})=\frac{3\gamma}{4\pi^2}[e_1(E)^2+e_2(E)^2+e_1(E^{'})^2+e_2(E^{'})^2+\gamma^2]^{-5/2},
\end{equation}
where the distribution width  $\gamma$~=~$\frac{\pi}{27}$$\frac{1+\sigma}{1-\sigma}$$\{\sum_i[c[i]\Delta$\textit{V}$[i]/V)^2\}^{1/2}$ depends on the Poisson ratio of diamond $\sigma$~=~0.2,\cite{field12} the fractional concentration $c[i]$ of the $i^{th}$ species of point defect and the volume strain $\Delta V[i]/V$ per unit fractional concentration of defects of the $i^{th}$ species.\cite{Klimin_Malkin_2010,Malkin_PRB_2012} The strain broadened lineshape of the IR ZPL is then the convolution of \textit{g}(\textbf{e}) and a Gaussian lineshape
\begin{equation}
\label{convolution}
\begin{array}{c}
 \displaystyle
I(\nu)\propto\int d\textbf{e}g(e)(exp\left\{-\frac{[\nu-\nu_0-\nu_-]^2}{2\Delta^2}\right\}+
\\
\\
 \displaystyle
exp\left\{-\frac{[\nu-\nu_0-\nu_+]^2}{2\Delta^2}\right\})
\:,
\end{array}
\end{equation}
where the Gaussian width $\Delta$ accounts for the broadening induced by the deformations of A$_1$ symmetry, all other inhomogeneous broadening (not arising from strain induced by point defects) and homogeneous broadening. In the above, the integration is taken over the four E symmetry deformation tensor components, and $\nu_0$ is the energy of the unperturbed IR ZPL. This integration must be performed numerically. Note that we have not explicitly treated the broadening induced by the deformations of A$_1$ symmetry because, as supported by the following observations, the smaller strain shift of the IR ZPL implies that their contributions to the width $\Delta$ are negligible compared to the other sources of inhomogeneous and homogeneous broadening.

\begin{figure}
\begin{center}
    \includegraphics[width=0.95\columnwidth]{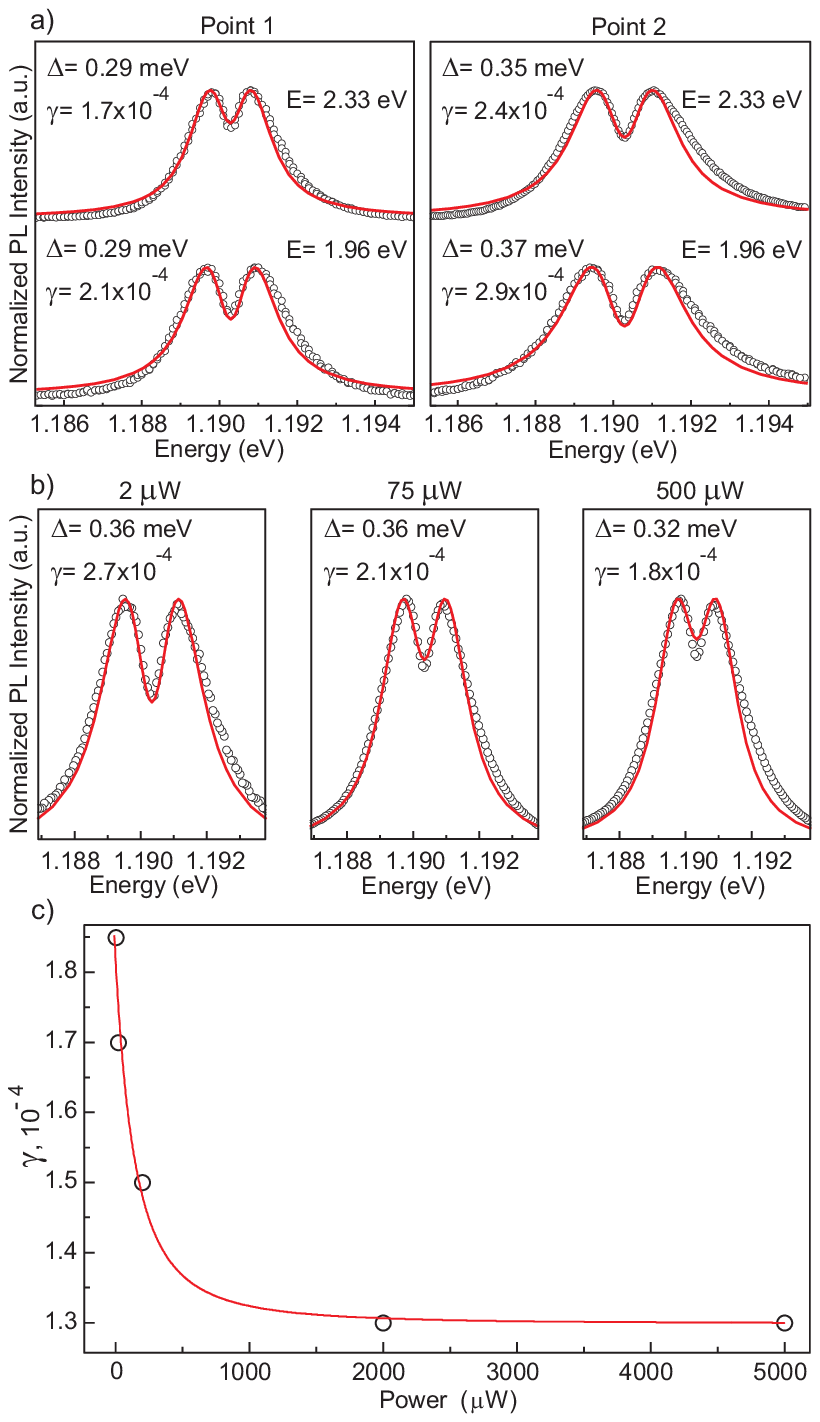}
\caption{\label{fig3} Experimental (circles) and simulated (solid lines) IR ZPL spectra (a) for different excitation energies (2.33~eV and 1.96~eV) at two points of S2 taken at the laser power P~=~70~$\mu$W, and (b) for different powers of the 2.33~eV excitation  laser. (c) the power dependence of the extracted value of $\gamma$ (points) from spectra at a third point in S2 with 2.33~eV excitation. The solid line denotes the fit described in the text. All spectra were recorded at temperature \textit{T}~=~78~K.}
\end{center}
\end{figure}

As demonstrated in Fig. 2(a), our expression for $I(\nu)$ successfully describes the inhomogeneous lineshape of the IR ZPL in S2. Fig. 2(b) depicts the temperature dependence of the IR ZPL under the same photoexcitation conditions. The dashed lines in Fig. 2(b) indicate that the ZPL splitting does not change with temperature, thereby implying that the strain induced broadening is temperature independent. The temperature dependence of the ZPL rather arises from the $\propto T^3$ temperature variation of the Gaussian width $\Delta$ that is plotted in Fig. 2(c). Such temperature dependent broadening has been observed in the NV$^-$ center's visible ZPL\cite{Fu_PRL_2009} and the ZPLs of other defects in diamond,\cite{jahnke15} and attributed to temperature dependent electron-phonon scattering. This is likely to be the case also for the IR ZPL, however it must be noted that some caution must be exercised because the Gaussian width $\Delta$ is not purely related to the homogeneous broadening that would arise from electron-phonon scattering, but also has contributions from other mechanisms of inhomogeneous broadening. If this complication is ignored, then our results suggest that the electron-phonon scattering affecting the IR ZPL has a somewhat unusual $\propto T^3$ temperature variation. Although a similar temperature variation has been observed for the ZPL of the negatively charged silicon-vacancy (SiV$^-$) center, this has been shown to be a modification of the usual $\propto T^5$ Raman-type scattering mechanism by the large spin-orbit interaction of the SiV$^-$ center.\cite{jahnke15} No such spin-orbit interaction can be present in the singlet levels involved in the NV$^-$ center's IR ZPL. It is possible that the $\propto T^3$ temperature variation of the IR ZPL arises from the presence of a strong Jahn-Teller effect,\cite{Hizhnyakov04} which is consistent with the dramatic asymmetry of photoabsorption and PL spectra of the IR ZPL's vibrational sideband that is currently poorly understood.\cite{kehayias13} However, substantial further investigation is required before this conclusion can be drawn.

To test the role of N$^0$ in strain broadening, we analysed the dependence of the IR ZPL on the photoexcitation energy and power. As can be seen in Fig. 3(a), increasing the photoexcitation energy (decreasing wavelength) with fixed power and temperature results in the decrease of $\gamma$ with almost constant $\Delta$. A result that is consistent over the two different positions in S2 that we sampled. Fig. 3(b) demonstrates that increasing the photoexcitation power with fixed energy and temperature similarly results in the decrease of $\gamma$ with almost constant $\Delta$. These results support the hypothesis described above that increased photoionization of N$^0$ at higher photoexcitation energies and powers will reduce strain broadening. This hypothesis is further supported by the fit of the photoexcitation power dependence of $\gamma$ in Fig. 3(c) by a simple two-state rate equation model of N$^0$ photoionization. Defining a power dependent photoionization rate $BP$ and a power independent recombination rate $A$, the steady state concentration of N$^0$ is
\begin{equation}
c[\mathrm{N}^0](P) = c[\mathrm{N}^0](0)\frac{1}{1+\Omega P}
\end{equation}
where $c[\mathrm{N}^0](0)$ is the concentration of N$^0$ at zero power and $\Omega=B/A$. The steady state expression for $\gamma$ is then
\begin{equation}
\gamma(P) = \{\gamma[\mathrm{other}]^2+[\gamma[\mathrm{N}^0](0)\frac{1}{1+\Omega P}]^2\}^{1/2}
\end{equation}
where $\gamma[\mathrm{N}^0](0)=\frac{\pi}{27}\frac{1+\sigma}{1-\sigma}c[\mathrm{N}^0](0)\frac{\Delta V[\mathrm{N}^0]}{V}$ is the strain broadening due to N$^0$ at zero power and $\gamma[\mathrm{other}]$ is the power independent strain broadening due to all other sources. The parameter values obtained from the fit in Fig. 3(c) are $\gamma[\mathrm{other}]=1.30\pm0.04$, $\gamma[\mathrm{N}^0](0)=1.27\pm0.07$ and $\Omega=0.004\pm0.002$ $\mathrm{\mu W}^{-1}$.

The power dependence of the strain broadening $\gamma(P)$ appears to imply the unexpected conclusion that N$^0$ causes at most $\gamma[\mathrm{N}^0](0)/\{\gamma[\mathrm{other}]+\gamma[\mathrm{N}^0](0)\}\approx50\%$ of the strain broadening, thereby indicating that major sources of strain broadening are yet to be identified. The other defects known to be in the sample are NV$^-$, its neutral charge state NV$^0$, and the vacancy (V) center. However, given that the concentrations of NV$^-$ and NV$^0$ must be less than N$^0$ by growth arguments, and that our annealing of S2 is expected to almost completely remove V centers, for these centers to be the major sources of strain broadening, they must induce much larger strains than N$^0$.

Another issue arises in the quantitative analysis of the fit parameters. Using the value of $\Delta V[\mathrm{N}^0]/V=0.41$ that has been obtained from x-ray crystallography,\cite{lang91} the fitted value of $\gamma[\mathrm{N}^0](0)$ yields a N$^0$ concentration of $c[\mathrm{N}^0](0)\approx 1775$ ppm at zero photoexcitation power, which is $\sim10$ times larger than the specification of $\lesssim200$ ppm that was assigned by the sample supplier. Possible origins of this overestimation are incorrect specification by the supplier, large non-uniformity of N$^0$ concentration over the sample or photoionization of NV$^-$. Considering the latter further, it is known that NV$^-$ photoionizes to NV$^0$ under excitation with energy $>1.946$ eV.\cite{Doherty_PhysRep_2013} Thus, photoionization of NV$^-$ could occur under the excitation conditions we employed. However, similar to before, for photoionization of NV$^-$ to explain the overestimation, the difference in strain induced by NV$^-$ and NV$^0$ must be much larger than the difference between N$^0$ and N$^+$.

To test these explanations, we have performed density functional theory (DFT) calculations of the strain induced by N$^0$, N$^+$, NV$^-$ and NV$^0$ centers. Calculations were performed using the VASP program\cite{VASP} and employed the screened hybrid functional of Heyd, Scuseria, and Ernzerhof HSE06 to describe the electronic structure. Interaction between ions and valence electrons were treated with the PAW approach.\cite{VASP} Wavefunctions were expanded in plane waves (using 400 eV for the kinetic energy cutoff). We have used 216 (512) atom supercells to model N$^0$ and N$^+$ (NV$^-$ and NV$^0$).  The Brillouin zone was sampled at the $\Gamma$ point to make sure local symmetry is properly represented. Defect-induced stress was calculated as the average of the three diagonal stress components $p=(\sigma_{xx}+\sigma_{yy}+\sigma_{yy})/3$ in the supercell, mimicking the averaging over different defect orientations.
Subsequently, $p$ was corrected for a small residual stress of the pristine supercell when calculated using exactly the same computational parameters (k-point sampling, cutoff energy, and the supercell size). Crucially, stress of charged defects (N$^+$ and NV$^-$) has been corrected for the spurious potential alignment term contribution using the procedure of Bruneval et al [Eq.~(26) in the Ref. \onlinecite{Bruneval_PRB_2015}]. Absolute deformation potentials of bulk diamond states needed for that procedure have been taken from calculations reported in Ref.~\onlinecite{Li_PRB_2006}. Finally, volume expansion coefficient has been determined via $V[i]/V=p/(Bx)$, where $B=481$ GPa is the calculated bulk modulus of diamond (cf.~the experimental value of 443 GPa), and $x$ is the concentration of defects in our supercells (either $1/216$ or $1/512$).

\begin{table}
\caption{The volume strain per unit concentration of various defects in diamond as we have calculated using DFT and as obtained previously\cite{lang91} by experiment.}
\begin{ruledtabular}
\begin{tabular}{lllll}
 $V[i]/V$ &  N$^0$ & N$^+$ & NV$^-$ & NV$^0$  \\
 \hline
Calc. & 0.23 & 0.01 & $-0.16$ & 0.00 \\
Expt. & 0.41 & -- & -- & -- \\
\end{tabular}
\end{ruledtabular}
\end{table}

The results of our DFT calculations are summarized in Table 1. The value for the volume strain $(V[\mathrm{N}^0]/V)$ per unit fractional concentration of N$^0$ underestimates the experimental value by $\sim$$50\%$.
%which indicates the absolute accuracy of the calculations.
%
Part of this disagreement stems from the shortcomings of DFT and, likely, relatively large defect concentrations in actual calculations. However, the disagreement can also arise from experimental errors in determining the actual nitrogen concentration.
In agreement with expectations, the value for N$^+$ is negligible. The value for NV$^-$ is negative, which is consistent with the expectations based upon the presence of a vacancy in the defect.
Interestingly, the value of $V[i]/V$ for NV$^0$ is almost zero, which reflects a small outward relaxation of atoms adjacent to the vacancy in NV$^0$ as compared to NV$^-$.
Importantly, the magnitudes of the values for N$^0$ and NV$^-$ are comparable and thus, in accordance with the arguments above, we can conclude that NV$^-$ is not the unidentified major source of broadening nor is its photoionization the explanation for the overestimation of the N$^0$ concentration. The resolution of these issues requires further investigation beyond this work. Such future work should include the confirmation of the average concentration of N$^0$ in the samples using electron paramagnetic resonance and infrared absorption spectroscopy as well as the systematic study of large non-uniformities of the N$^0$ concentration in the samples. Indeed, evidence of the latter can be immediately seen in the variation of $\gamma$ between the two positions in S2 that are depicted in Fig 3. If the issues can be resolved, then analysis of the strain broadening of the NV$^-$ center's IR ZPL will be a promising tool for defect micro-/nanoscale tomography.

\section{\label{sec:level1}Conclusion}

To summarize, we have reported a detailed study of the splitting and inhomogeneous broadening of the NV$^-$ center's IR ZPL. We have identified that its inhomogeneous lineshape differs from the center's visible ZPL because the IR ZPL has a different susceptibility to strain fields and is not susceptible to electric fields. This singular susceptibility motivated the use of the IR ZPL to quantify the broadening due to different microscopic distributions of defects via the strain fields that they generate. We developed a model of the inhomogeneous lineshape of the IR ZPL by adapting the strain broadening theory proposed in Refs \onlinecite{Klimin_Malkin_2010} and \onlinecite{Malkin_PRB_2012}. We applied the model to interpret observations of temperature variation of the IR ZPL shape. By distinguishing strain broadening from other sources of broadening, we identified an anomalous $\propto T^3$ temperature dependent broadening mechanism, which potentially provides additional evidence of a strong Jahn-Teller effect in the $^1$E level. This additional evidence furthers the understanding the singlet levels of the NV$^-$ center and their role in the center's optical spin polarization and readout mechanisms.

We interestingly found that the strain broadening of the IR ZPL can be reduced by increasing excitation intensity or reducing excitation wavelength. Through the application of our model, these effects were found to be qualitatively consistent with the photodependence of the concentration of N$^0$ centers. However, quantitative analysis revealed that only $\approx50\%$ of the strain broadening can be attributed to the photodependence of N$^0$, thereby indicating that there exist other, yet to be identified, sources of strain broadening. Furthermore, the quantitative analysis yielded an estimate of the local N$^0$ concentration that was $\sim10$ times larger than what the supplier specified as the average concentration. To investigate these issues further, we performed DFT calculations of the strain induced by different charge states of N and NV centers. These calculations supported our initial interpretation and thus did not resolve these issues of the quantitative analysis. We conclude that further work is required to identify the unidentified sources of broadening and to test the estimation of the local concentration of N$^0$. Given this further work, we proposed that the IR ZPL has the potential to be used as a sensitive spectroscopic tool to probe the microscopic distribution of strain fields and to perform tomography of defect concentrations.

\begin{acknowledgments}
This work has been supported by the ARC grants FT110100225, CE140100003 and DE170102735, and performed according to the Russian Government Program of Competitive Growth of Kazan Federal University.  B. C. G acknowledges the support of the ARC Centre of Excellence for Nanoscale BioPhotonics (CNBP) and an ARC Future Fellowship. V. A. S. acknowledges the support of the RFBR Project 16-52-76017 (ERA.Net RUS Plus project ”DIABASE”). A. A. acknowledges the support of the Research Council of Lithuania via Grant No. M-ERA.NET-1/2015. Computations were performed at the High Performance Computing Center ``HPC Sauletekis'' in Faculty of Physics, Vilnius University.  We wish to acknowledge fruitful discussions with Boris Malkin, Sergey Feofilov and Neil Manson.
\end{acknowledgments}


\begin{thebibliography}{}

\bibitem{qs1}
G.R. Balasubramanian, P. Neumann, D. Twitchen, M. Markham, R. Kolesov, N. Mizuochi, J. Isoya, J. Achard, J. Beck, J. Tissler, V. Jacques, P.R. Hemmer, F. Jelezko and J. Wrachtrup, \textit{Ultralong spin coherence time in isotopically
engineered diamond}, Nature Mat. \textbf{8}, 383 (2009).

\bibitem{qs2}
K. Arai, C. Belthangady, H. Zhang, N. Bar-Gill, S.J. DeVience, P. Cappellaro, A. Yacoby and R.L. Walsworth, \textit{Fourier magnetic imaging with nanoscale resolution and compressed sensing speed-up using electronic spins in diamond}, Nature Nanotech. \textbf{10}, 589 (2015).

\bibitem{qs3}
F. Dolde, H. Fedder, M.W. Doherty, T. Nöbauer, F. Rempp, G. Balasubramanian, T. Wolf, F. Reinhard, L.C.L. Hollenberg, F. Jelezko and J. Wrachtrup, \textit{Electric-field sensing using single diamond spins}, Nat. Phys. \textbf{7}, 459 (2011).

\bibitem{qs4}
G. Kucsko, P.C. Maurer, N.Y. Yao, M. Kubo, H.J. Noh, P. K. Lo, H. Park and M.D. Lukin, \textit{Nanometre-scale thermometry in a living cell} Nature {\bf 500}, 54 (2013).

\bibitem{qs5}
M.W. Doherty, V.V. Struzhkin, D.A. Simpson, L.P. McGuinness, Y. Meng, A. Stacey, T.J. Karle, R.J. Hemley, N.B. Manson, L.C.L. Hollenberg and S. Prawer, \textit{Electronic properties and metrology applications of the diamond NV−  center under pressure}, Phys. Rev. Lett. {\bf 112}, 047601 (2014).

\bibitem{qs6}
D. R. Glenn, K. Lee, H. Park, R. Weissleder, A. Yacoby, M.D. Lukin, H. Lee, R.L. Walsworth and C.B. Connolly, \textit{Single-cell magnetic imaging using a quantum diamond microscope}, Nature methods {\bf 12}, 8 (2015).

\bibitem{maser}
L. Jin, M. Pfender, N. Aslam, P. Neumann, S. Yang, J. Wrachtrup and R.-B. Liu, \textit{Proposal for a room-temperature diamond maser}, Nat. Comm. {\bf 6}, 8251 (2015).

\bibitem{qc1}
G. Waldherr, Y. Wang, S. Zaiser, M. Jamali, T. Schulte-Herbrueggen, H. Abe, T. Ohshima, J. Isoya, J.F. Du, P. Neumann and J. Wrachtrup, \textit{Quantum error correction in a solid-state hybrid spin register}, Nature  \textbf{506}, 204 (2014).

\bibitem{qc2}
N.Y. Yao, L. Jiang, A.V. Gorshkov, P.C. Maurer, G. Giedke, J.I. Cirac and M.D. Lukin, \textit{Scalable architecture for a room temperature solid-state quantum information processor}, Nature Comm. \textbf{3}, doi:10.1038/ncomms1788 (2012).

\bibitem{qc3}
B. Hensen, H. Bernien, A.E. Dréau, A. Reiserer, N. Kalb, M.S. Blok, J. Ruitenberg, R.F.L. Vermeulen, R.N. Schouten, C. Abell\'{a}n, W. Amaya, V. Pruneri, M.W. Mitchell, M. Markham, D.J. Twitchen, D. Elkouss, S. Wehner, T.H. Taminiau and R. Hanson, \textit{Loophole-free Bell inequality violation using electron spins separated by 1.3 kilometres}, Nature \textbf{526}, 682 (2015).

\bibitem{stoneham69}
A.M. Stoneham, \textit{Shapes of inhomogenously broadened resonance lines in solids}, Rev. Mod. Phys. \textbf{41}, 82 (1969).

\bibitem{davies70}
G. Davies, \textit{No phonon lineshape and crystal strain fields in diamonds}, J.Phys. C: Solid St. Phys \textbf{3}, 2474 (1970).

\bibitem{davies71}
G. Davies, \textit{Approximate widths of zero phonon lines broadened by point defect strain fields}, J. Phys. D: Appl. Phys. \textbf{4}, 1340 (1971).

\bibitem{Doherty_PhysRep_2013}
M.W. Doherty, N. B. Manson, P. Delaney, F. Jelezko, J. Wrachtrup and L.C.L. Hollenberg, \textit{The nitrogen-vacancy colour centre in diamond}, Phys. Rep. \textbf{528}, 1 (2013).

\bibitem{goldman1}
M.L. Goldman, A. Sipahigil, M.W. Doherty, N.Y. Yao, S.D. Bennett, M. Markham, D.J. Twitchen, N.B. Manson, A. Kubanek and M.D. Lukin, \textit{Phonon-induced population dynamics and intersystem crossing in nitrogen-vacancy centers}, Phys. Rev. Lett. \textbf{114}, 145502 (2015).

\bibitem{goldman2}
M.L. Goldman, M.W. Doherty, A. Sipahigil, N.Y. Yao, S.D. Bennett, N.B. Manson, A. Kubanek and M.D. Lukin, \textit{State-selective intersystem crossing in nitrogen-vacancy centers}, Phys. Rev. B \textbf{91}, 165201 (2015).

\bibitem{Mita_1996}
Y. Mita, \textit{Change of absorption spectra in type-Ib diamond with heavy neutron irradiation}, Phys. Rev. B {\bf 53}, 11360 (1996).

\bibitem{Budker_PRB_NV_production_2010}
V.M. Acosta, E. Bauch, M.P. Ledbetter, C. Santori, K.-M. C. Fu, P.E. Barclay, R.G. Beausoleil, H. Linget, J.F. Roch, F. Treussart, S. Chemerisov, W. Gawlik and D. Budker, \textit{Diamonds with a high density of nitrogen-vacancy centers for magnetometry applications},  Phys. Rev. B {\bf 80}, 115202 (2010).

\bibitem{Rogers_Doherty_2015}
L.J. Rogers, M.W. Doherty, M.S.J. Barson, S. Onoda, T. Ohshima and N.B. Manson, \textit{Singlet levels of the NV-  centre in diamond}, New J. Phys. {\bf 17}, 013048 (2015).

\bibitem{Doherty_NJP_2011}
M.W. Doherty, N.B. Manson, P. Delaney and L.C.L. Hollenberg, \textit{The negatively charged nitrogen-vacancy centre in diamond: the electronic solution}, New J. Phys. {\bf 13}, 025019 (2011).

\bibitem{stoneham92}
A.M. Stoneham, \textit{Theoretical status of diamond and its defects, excited states and atomic motion}, Materials Science and Engineering \textbf{B11}, 211 (1992).

\bibitem{Loubser_vanWyk_1978}
J.H.N. Loubser and J.A. van Wyk, \textit{Electron spin resonance in the study of diamond}, Rep. Prog. Phys. \textbf{41}, 1201 (1978).

\bibitem{Farrer_1969}
R.G. Farrer, \textit{On the substitutional nitrogen donor in diamond} Sol. St. Comm. {\bf 7}, 685 (1969).

\bibitem{Isberg_PRB_2006}
J.Isberg, A.Tajani and D.J. Twitchen, \textit{Photoionization measurement of deep defects in single-crystalline CVD diamond using the transient-current technique}, Phys. Rev. B {\bf 73}, 245207 (2006).

\bibitem{Klimin_Malkin_2010}
S.A. Klimin, D. S. Pytalev, M. N. Popova, B. Z. Malkin, M. V. Vanyunin and S. L. Korableva, \textit{High-resolution optical spectroscopy of Tm3+ ions in LiYF4: Crystal-field energies, hyperfine and deformation splittings, and the isotopic structure}, Phys. Rev. B {\bf 81}, 045113 (2010).

\bibitem{Malkin_PRB_2012}
B.Z. Malkin, D.S. Pytalev, M.N. Popova, E.I. Baibekov, M.L. Falin, K.I. Gerasimov and N.M. Khaidukov, \textit{Random lattice deformations in rare-earth-doped cubic hexafluoroelpasolites}, Phys. Rev. B {\bf 86}, 134110 (2012).

\bibitem{field12}
J.E. Field, \textit{The mechanical and strength properties of diamond} Rep. Prog. Phys. \textbf{75}, 126505 (2012).

\bibitem{Fu_PRL_2009}
K.M.C. Fu, C. Santori, P.E. Barclay, L.J. Rogers, N.B. Manson and R.G. Beausoleilet, \textit{Observation of the dynamic Jahn-Teller effect in the excited states of Nitrogen-Vacancy Centers in diamond}, Phys. Rev. Lett. {\bf 103}, 256404 (2009).

\bibitem{jahnke15}
K.D. Jahnke, A. Sipahigil, J.M. Binder, M.W. Doherty, M. Metsch, L.J. Rogers, N.B. Manson, M.D. Lukin and F. Jelezko, \textit{Electron–phonon processes of the silicon-vacancy centre in diamond}, New J. Phys. \textbf{17}, 043011 (2015).

\bibitem{Hizhnyakov04}
V. Hizhnyakov, V. Boltrushko, H. Kaasik and I. Sildos, \textit{Phase relaxation in the vicinity of the dynamic instability: anomalous temperature dependence of zero-phonon line}, J. Lumin. \textbf{107}, 351 (2004).

\bibitem{kehayias13}
P. Kehayias, M.W. Doherty, D. English, R. Fischer, A. Jarmola, K. Jensen, N. Leefer, P. Hemmer, N.B. Manson and D. Budker, \textit{Infrared absorption band and vibronic structure of the nitrogen-vacancy center in diamond}, Phys. Rev. B \textbf{88}, 165202 (2013).

\bibitem{lang91}
A.R. Lang, M. Moore, A.P.W. Makepeave, W. Wierzchowski and C.M. Welbourn, \textit{On the dilation of synthetic type Ib diamond by substitutional nitrogen impurity}, Phil. Trans. R. Soc. Lond. A \textbf{337}, 497 (1991).

\bibitem{VASP}
G. Kresse and J. Furthm\"{u}ller,
\textit{Efficient iterative schemes for ab initio total-energy calculations using a plane-wave basis set},
Phys. Rev. B \textbf{54}, 11169 (1996);
G. Kresse and D. Joubert,
\textit{From ultrasoft pseudopotentials to the projector augmented-wave method},
Phys. Rev. B {\bf 59}, 1758 (1999).

\bibitem{Heyd_JCP_2003}
J. Heyd, G. E. Scuseria, and M. Ernzerhof,
\textit{Hybrid functionals based on a screened Coulomb potential},
J. Chem. Phys. {\bf 118}, 8207 (2003).

\bibitem{Bruneval_PRB_2015}
F. Bruneval, C. Varvenne, J.-P. Crocombette, and E. Clouet,
\textit{Pressure, relaxation volume, and elastic interactions in charged simulation cells},
Phys. Rev. B {\bf 91}, 024107 (2015).

\bibitem{Li_PRB_2006}
Y.-H. Li, X. G. Gong, and S.-H. Wei,
\textit{Ab initio all-electron calculation of absolute volume deformation potentials of IV-IV, III-V, and II-VI semiconductors: The chemical trends},
Phys. Rev. B {\bf 73}, 245206 (2006).

\end{thebibliography}
\end{document}